\documentclass[12pt,preprint]{aastex}

\begin{document}


\title{Substructure in Tidal Streams; Tributaries in the Anticenter Ring}

\author{C. J. Grillmair}
\affil{Spitzer Science Center, 1200 E. California Blvd., Pasadena,  CA 91125}
\email{carl@ipac.caltech.edu}

\begin{abstract}

We report on the detection in Sloan Digital Sky Survey data of at
least three, roughly parallel components in a $65\arcdeg$-long,
stellar stream complex previously identified with the Anticenter or
Monoceros Ring.  The three-stream complex varies in width from
$4\arcdeg$ to $6\arcdeg$ along its length and appears to be made up of
two or more narrow substreams as well as a broader, diffuse
component. The width and complexity of the stream indicate that the
progenitor was likely a dwarf galaxy of significant size and mass. The
stream is 8.9 kpc distant and is oriented almost perpendicularly to
our line of sight. The visible portion of the stream does not pass
near any known dwarf galaxies and a preliminary orbit does not point
to any viable progenitor candidates. Orbits for the narrower
substreams can be modeled with velocity offsets from the broad
component of $\approx 8$ km s$^{-1}$. We suggest that the broad
component is likely to be the remains of a dwarf galaxy, while the
narrower streams constitute the remnants of dynamically distinct
components which may have included a native population of globular
clusters. While the color of the main sequence turn-off is not unlike
that for the Monoceros Ring, neither the visible stream nor any
reasonable projection of its orbit passes through Monoceros or Canis
Major, and we conclude that this stream is probably unrelated to the
overdensities found in these regions.

\end{abstract}



\section{Introduction}

The value of large scale digital sky surveys to studies of Galactic
structure and the Local Group has become abundantly clear in recent
years, and particularly in the last few months.  In addition to the
large scale features attributed to past galaxy accretion events
\citep{yann03,maje2003,roch04}, Sloan Digital Sky Survey (SDSS) data
were used to detect the remarkably strong tidal tails of Palomar 5
\citep{oden2001,rock2002, oden2003, grill2006b}, NGC 5466
\citep{belo2006a, grill2006a}, and streams due to extant or extinct
globular clusters \citep{grill2006c} and dwarf galaxies
\citep{belo2006b, grill2006d}. \citet{will2005}, \citet{zuck2006}, and
\citet{belo2006c} recently used SDSS data to discover several new dwarf
satellites of the Milky Way.

In this paper we continue our analysis of the SDSS database to search
for extended structures in the Galactic halo. We briefly describe our
analysis in Section \ref{analysis}. We discuss the structure of a
stream complex, attributed to the Monoceros Ring by \citet{belo2006b}
in Section \ref{discussion}, estimate distances in Section
\ref{distance}, and put constraints on the orbit in Section
\ref{orbit}. 

\section{Data Analysis \label{analysis}}

Data comprising $ugriz$ photometry for stars in the region $108\arcdeg
< \alpha < 162\arcdeg$ and $-4\arcdeg < \delta < 68\arcdeg$ were
extracted from Data Release 5 of the SDSS using the SDSS CasJobs query
system. The data were analyzed using the matched filter technique
employed by \citet{grill2006a}, \citet{grill2006b},
\citet{grill2006c}, \citet{grill2006d} and described in detail by
\citet{rock2002}. This technique is made necessary by the fact that,
over the magnitude range and over the region of sky we are
considering, the foreground disk stars outnumber the more distant
stars in the Galactic halo by some three orders of magnitude. Applied in
the color-magnitude domain, the matched filter is a means by which we
can optimally differentiate between two populations.

We used the SDSS photometry to create a color-magnitude density or
Hess diagram for both stars of interest and for the foreground
population. Dividing the former by the latter, we generated an array
of relative weights which we used as an optimal color-magnitude
filter.  We generated the search filter using the color-magnitude
distribution of stars in M 13. Owing to M 13's proximity, we have a
better measure of its luminosity function and the effects of SDSS
completeness than we do for other objects in the DR5 field.  We used
all stars with $15 < g < 22.5$. We dereddended the SDSS photometry as
a function of position on the sky using the DIRBE/IRAS dust maps of
\citet{schleg98}. A single Hess diagram for field stars was
generated using $1.2 \times 10^7$ stars spread over $\sim 2200$
deg$^2$ of DR5.  We applied the M 13 filter to the entire survey area,
and the resulting weighted star counts were summed by location on the
sky to produce a two dimensional probability map.

In Figure 1 we show the final filtered star count distribution, using
a filter matched to the $g, g - i$ color magnitude distribution of
stars in M 13 but shifted faintwards by 0.3 mags. An exponential
surface function has been subtracted from the image to remove the
rapid rise in the number of disk stars at low Galactic latitudes, and
a 4th-order polynomial surface has been used to remove remaining large
scale gradients. The final image has been binned to a pixel size of
$0.08\arcdeg$ and smoothed using a Gaussian kernel with $\sigma =
0.2\arcdeg$.

\section{Discussion \label{discussion}}

Apparent in Figure 1 is a broad and complex stream running from north
to south across the field. Portions of this stream are visible in
\citet{newberg02} and \citet{belo2006b}. The narrow, curved stream
running towards the northeast is part of the 63$\arcdeg$-long globular
cluster stream found by \citet{grill2006c}, hereafter referred to as
GD-1.  The broad, east-west stream just above the main gap in the data
is the Sagittarius stream discussed by \citet{belo2006b}, though
somewhat muted by our filtering due to its greater distance.  The
stream complex of interest extends from (R.A., decl.) =
($126.4\arcdeg, -0.7\arcdeg$) to (R.A., decl.) = ($133.9\arcdeg,
64.2\arcdeg$), and runs in a $65\arcdeg$, nearly great circle path
from Ursa Major in the north to Hydra in the south. The stream is
truncated at both the southern and northern ends by the limits of the
available data. We note that there is also an apparent concentration
of stars at (R.A., decl.) = ($134\arcdeg, 3.4\arcdeg$), surrounded by
faint, banded substructure roughly parallel to that of the main
western stream.

From the reddening map of \citet{schleg98}, the maximum values of
$E(B-V)$ are $\approx 0.2$ (near the northern tip), with typical
values near 0.03 along the remainder of the stream.  There are
diminutions here and there in the stream that could be attributed to
regions of higher reddening, but there are no long features with a
north-south orientation which could be held to account for either the
appearance of the stream as a whole, or for the different components
within it.

Sampling at several representative points, the main stream complex
appears to be about $5\arcdeg$ wide on average. This is significantly
broader than the globular cluster streams found by \citet{oden2003,
grill2006a, grill2006b} and \citet{grill2006c}. $5\arcdeg$ corresponds
to about 800 pc at our estimated distance to the stream (see below),
which is much larger than the tidal diameters of globular clusters.
We conclude that the progenitor was considerably more extended than a
globular cluster and was most likely a dwarf galaxy.

The stream is clearly not just a broad swath of stars. Rather, it
appears to be made up of a $\sim 2\arcdeg$ wide broad component
running roughly along the center of the stream complex, and at least
two narrower, $1\arcdeg$-wide streams to the east and west of the
broad component. The fine structure in the stream complex is
illustrated in Figure 2, where we have made several east-west slices
across the complex at various declinations. Comparing Figures 1 and 2,
there are indications of additional structure within the broad
component, and still other, more tenuous parallel streams to the east
and west of the three major components.

Integrating the background subtracted, weighted star counts along the
stream over a width of $\approx 5\arcdeg$ we find that the total
number of stars in the discernible stream down to $g = 22.5$ is $9200
\pm 1500$. The mean surface density of stars in the southern portion
of the stream ($-1\arcdeg < \delta < 14\arcdeg$) is about 67 stars
deg$^{-2}$, which is roughly twice the 32 stars deg$^{-2}$ (corrected
for $cos(\delta)$) found in the northern section ($39\arcdeg < \delta
< 52\arcdeg$). The highest local surface densities exceed 200 stars
deg$^{-2}$. If we assume a globular cluster-like luminosity function
in the stream, then we can use the color transformation equations of
\citet{smith2002}, the deep M 4 luminosity function of
\citet{richer2002}, and the mass-luminosity relation of
\citet{baraffe97} to extrapolate to fainter magnitudes. Integrating
over $3 < M_V < 17$, we estimate a total number of stars in the
visible portion of the stream of $33,000 \pm 5400$, a total luminosity
of $5500 \pm 900 L_\odot$, and a total estimated mass of $9300 \pm
1500 M_\odot$.

\subsection{Color-Magnitude Distribution and Distance to the Stream \label{distance}}

In Figure 3 we show $g, g - i$ color-magnitude distributions for the
stream stars, extracted by generating a Hess diagrams of stars lying
along $2\arcdeg$-wide regions covering the western and eastern halves
of the stream complex and subtracting a similar field star
distribution sampled over $\approx 500$ deg$^2$ to the east and west
of the stream complex.  Despite the somewhat limited statistics, a
clear signature of the turn-off and main sequence is evident in the
stream population.  Moreover, the turn-off regions of the
distributions match the dereddened, shifted main sequence locus of M
13 fairly well. The turn-offs in the eastern and western halves of the
complex lie at dereddened $g - i = 0.27$ and $0.3$, respectively. A
similar diagram in $g - r$ yields dereddened turn-off colors of $g - r
= 0.23$ and 0.25, respectively, with the difference most likely a due
to $\sim 0.02$ mag measurement uncertainty. These estimates lie
between values of 0.26 and 0.22 measured by \citet{newberg02} for the
Monoceros Ring and the Sagittarius stream, respectively. They are also
much bluer than \citet{newberg02}'s estimate of dereddened $g - r =
0.40$ for thick disk stars.  Based solely on turn-off color, we can
therefore rule out association of the stream complex with thick
disk stars. However, within the uncertainties, the turn-off colors are
consistent with those of either the Monoceros Ring or the Sagittarius
stream.

Varying the magnitude shift applied to M 13's main sequence locus from
-1.0 to +3.0 mag, we measured the foreground-subtracted, mean surface
density of stream stars in the regions $-1\arcdeg < \delta <
9\arcdeg$, $17\arcdeg < \delta < 39\arcdeg$, and $39\arcdeg < \delta <
63\arcdeg$.  To avoid potential problems related to a difference in
age between M 13 and the stream stars, we used only the portion of the
filter with $19.5 < g < 22.5$, where the bright cutoff is 0.8 mag
below M 13's main sequence turn-off. Though this reduces the stream
contrast somewhat, it provides sufficient integrated signal-to-noise
to enable main sequence fitting.

Fitting Gaussians to the mean surface densities as a function of
magnitude shift (e.g. \citet{grill2006c}, we find that the highest
contrasts occur for shifts of +0.31, +0.29, and +0.37 mag for the
eastern half of the stream complex over the declination ranges given
above, respectively. For the western half, we find that the filter
response peaks at shifts of +0.34, +0.25, and +0.38 mag, respectively.
The magnitude shifts for the two halves are highly consistent with one
another, and we conclude that there is no significant distance offset
from one side of the complex to the other. Adopting a distance to M 13
of 7.7 kpc \citep{harris96} we find an average heliocentric distance
of $d = 8.9 \pm 0.2$ kpc. The stream is roughly perpendicular to our
line of sight and slightly curved about the Galactic center as
expected. Our distance estimate is in excellent agreement with the
$\approx 9$ kpc found by \citet{ibata03} in their WFS-0801 field,
which is situated on the western edge of the stream complex at $\alpha
= 120.5, \delta = 40.3$.

\subsection{Constraints on the Orbit \label{orbit}}

The visible portion of the stream complex spans the Galactic
anticenter direction and, projecting a great circle path, is inclined
by $35\arcdeg$ to the Galactic plane. Though we are currently limited
by a lack of velocity information, for a given model of the Galactic
potential the progenitor's orbit is actually fairly well constrained
by the observed distance and orientation of the stream. Using the
Galactic model of \citet{allen91} (which includes a disk, bulge, and
spherical halo, and which \citet{grill2006a, grill2006b} and
\citet{grill2006c} found to work reasonably well for NGC 5466, Pal 5,
and GD-1), we use a least squares method to fit both the orientation
on the sky and the distance measurements in Section \ref{distance}. In
addition to a number of normal points lying along the central
component of the stream, we chose as a velocity fiducial point a
position at the northern end of the stream at (R.A., decl) =
(125.463\arcdeg, 51.492\arcdeg).

If we allow the proper motions to be free ranging and uninteresting
parameters, the model which best fits the data predicts $v_{LSR} = -18
\pm 10$ km s$^{-1}$ at the fiducial point, where the uncertainty
corresponds to the 95\% confidence interval. A projection of this
orbit is labeled C in Figure 4. We note that the uncertainty is
primarily determined by the large lever arm over which it has been
possible to measure relative distances. The 95\% range in
$v_{LSR}$ in turn predicts a range in perigalactic and apogalactic
radii of $6.6 < R_p < 6.9$ kpc and $16.8 < R_a < 17.3$ kpc. Of course,
these ranges do not take into account uncertainties in the absolute
distance of the stream (which depends on the uncertainty in M 13's
distance) or of the validity of \citet{allen91}'s Galactic model.

Given the very similar distances estimated for the eastern and western
portions of the stream, it is highly unlikely that the complex could
be a superposition of multiple wraps around the Galaxy. The two
narrower streams (E and W in Figure 4) can be reasonably well modeled
by 0.18 $mas$ yr$^{-1}$ offsets in east-west proper motion. At the
distance of the stream complex this amounts to $\approx 8$ km
s$^{-1}$. The entire stream complex is therefore likely to be the
remains of a dwarf galaxy which contained distinct components spanning
a range of binding energies. The narrower streams might, for example,
be the remnants of the parent galaxy's globular cluster
population. Piecing together the original structure and evolution of
the stream's progenitor will require detailed N-body modeling.

Integrating orbits for parameter sets spanning the range above, we
find that, with the exception of the Sagittarius dE, there are no
known dwarf galaxies within $5\arcdeg$ of the projected orbit.  The
Sagittarius dE lies $3.1\arcdeg$ from the projection of the best-fit
orbit, but the orbital planes of the the Sagittarius dE and the new
stream are clearly distinct (Figure 4); we attribute this apparent
proximity to the expected confluence of orbit projections in the
direction of the Galactic center and not to any physical association
between them.

The orientation of the stream on the sky puts fairly strict limits on
the plane of the orbit. The visible portion of the stream passes
through Lynx, Cancer, and Hydra. We find no reasonable combination of
parameters that would place the southern projection of the stream in
either Monoceros or Canis Major. Nor does the visible extent of the
stream complex fit either the prograde or retrograde models of the
Monoceros stream computed by \citet{penarrubia05}. Thus, even while
the turn-off colors appear to be similar, we conclude on orbital grounds
that this stream complex is unlikely to be related to either the
Monoceros stream or the Canis Major overdensity.

\acknowledgments

Funding for the creation and distribution of the SDSS Archive has been
provided by the Alfred P. Sloan Foundation, the Participating
Institutions, the National Aeronautics and Space Administration, the
National Science Foundation, the U.S. Department of Energy, the
Japanese Monbukagakusho, and the Max Planck Society.

{\it Facilities:} \facility{Sloan}.

\clearpage



\begin{figure}
\epsscale{0.6}
\plotone{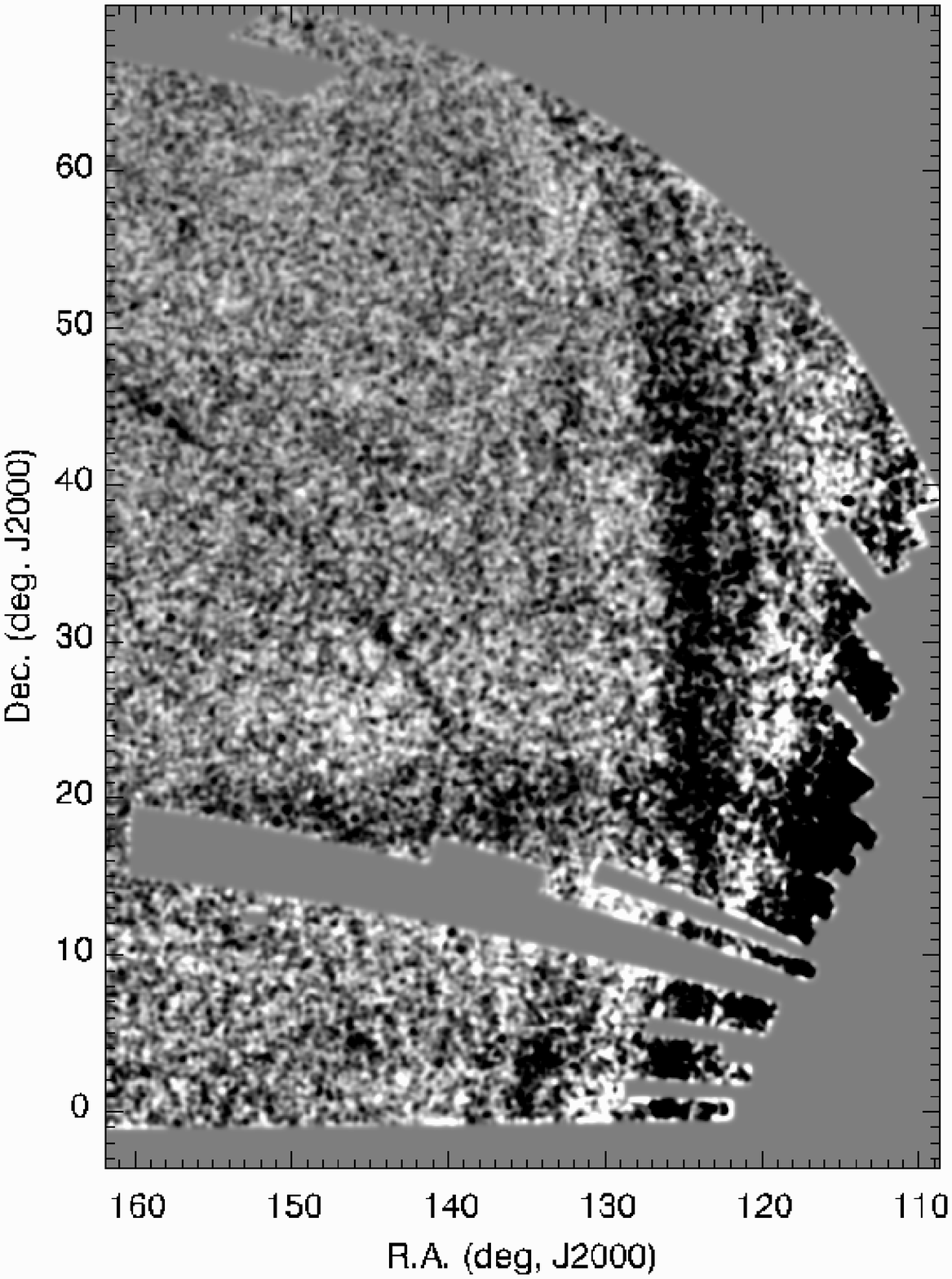}
\caption{Smoothed, summed weight image of the SDSS field after
subtraction of both an exponential and a 4th-order polynomial surface
fit. Darker areas indicate higher surface densities. The weight image
has been smoothed with a Gaussian kernel with $\sigma =
0.2\arcdeg$. The solid gray areas are either data missing from DR5, or
clusters or bright stars which have been masked out prior to
analysis. The stream complex runs from (R.A., decl) = ($126.4\arcdeg,
-0.7\arcdeg$) to approximately (R.A., decl.) = ($133.9\arcdeg,
64.2\arcdeg$). The feature at (R.A., decl.) = ($143\arcdeg,
30\arcdeg$) is a portion of the cold stellar stream discovered by
\citet{grill2006c}. \label{fig1}}
\end{figure}

\begin{figure}
\epsscale{0.7}
\plotone{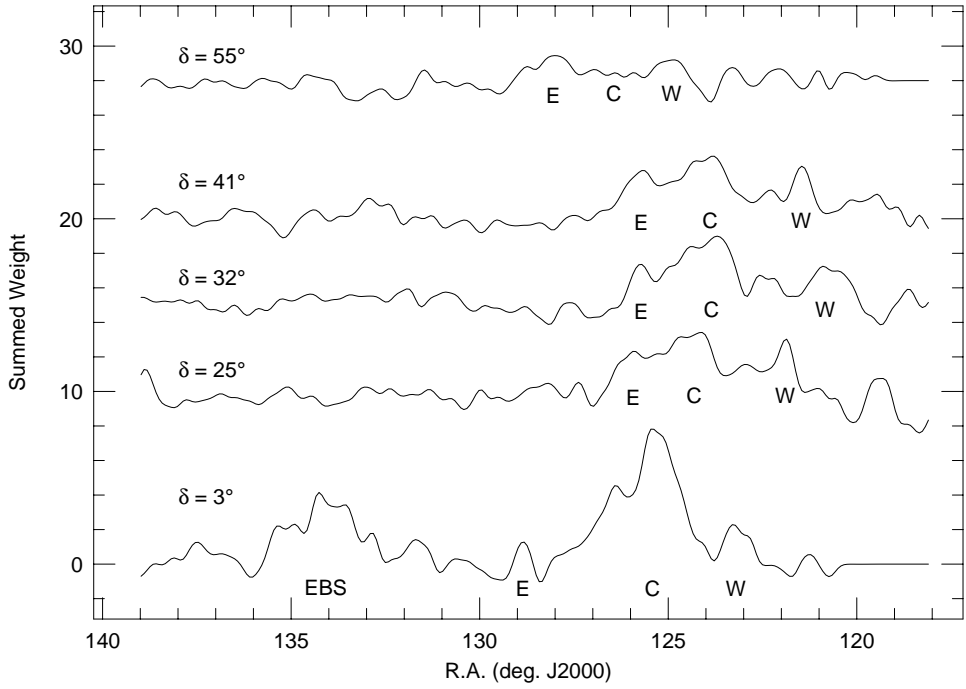}
\caption{Profiles across the stream complex at constant declination,
after smoothing the weighted star counts with a Gaussian kernel with
$\sigma = 0.25\arcdeg$ and summing over $2.4\arcdeg$-wide bands in
declination. E, C, and W indicate the Eastern, Central, and Western
components. EBS denotes the Easter Banded Structure alluded to in the
text. The profiles have been offset vertically for clarity. Note
indications of still other streams, both within the central stream and
to the east and west of the complex.}
\end{figure}

\begin{figure}
\epsscale{0.6}
\plotone{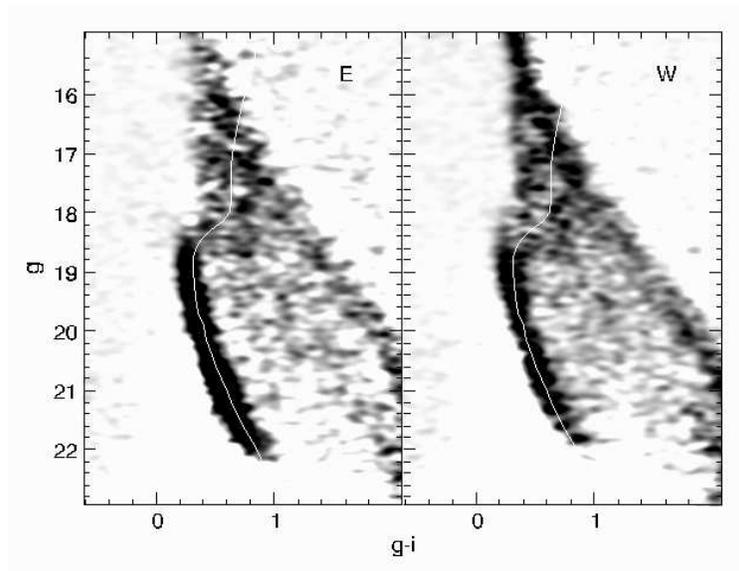}
\caption{The dereddened, background subtracted, color-magnitude
distribution of stream stars. The left-hand panel shows the C-M
distribution of stars in the western portion of the stream complex,
while the right-hand panel shows the distribution of stars in a
$2\arcdeg$-wide strip on the eastern side of the complex.  The main
sequence and the turn-off are clearly distinguishable. The solid line
shows the dereddened locus of giant branch and main sequence stars as
measured in DR5 for M 13, shifted faintwards by 0.3 mags.}
\end{figure}

\begin{figure}
\epsscale{0.6}
\plotone{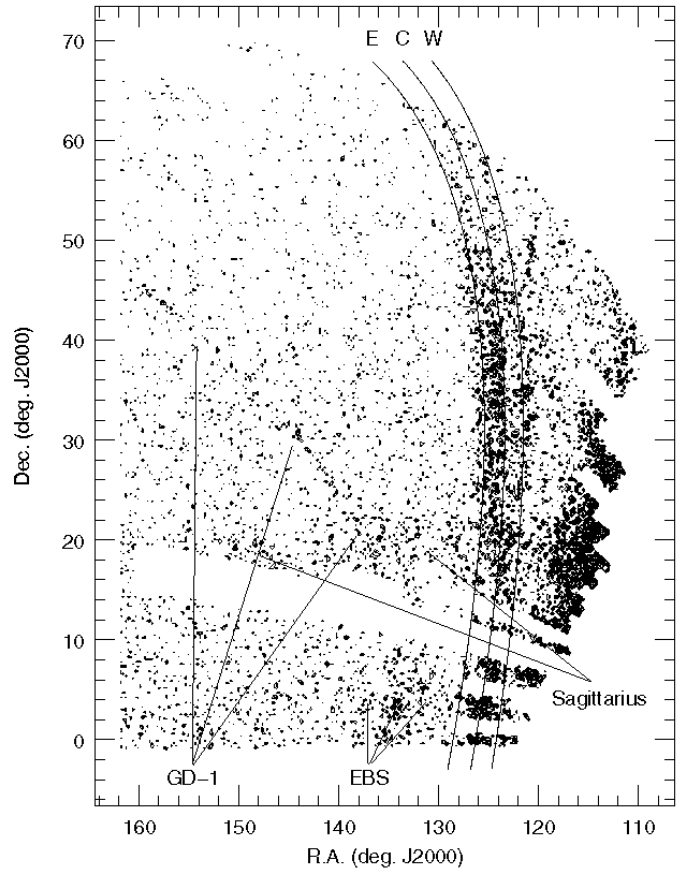}
\caption{Contour plot of Figure 1, showing projections of the best
fitting orbits to each of the eastern, central, and western stream
components. GD-1 is the globular cluster stream found by
\citet{grill2006c}, and other designations are as in Figure 2.}
\end{figure}

\end{document}